%% file: 00.tex
\documentclass{article}
\usepackage[utf8]{inputenc}

\input{00.sty}

\title{Python Type Hints are Turing Complete}
\author{Ori Roth}
\date{\today}

\begin{document}

\maketitle

\begin{abstract}
    \input{abstract}
\end{abstract}

\section{Introduction}
\label{section:aa}
\input{aa}

\paragraph*{Outline}
In \cref{section:python}, we apply Grigore's reduction to Python type hints and deduce that they are Turing complete.
\Cref{section:real} introduces an alternative design for subtyping machines that simulate the TMs in real time, i.e., make $O(1)$ operations for each TM transition.
Finally, \cref{section:implementation} presents our Python implementations of Grigore's original reduction and the new reduction and compares their performances.

\section{Python Subtyping is Undecidable}
\label{section:python}
\input{python}

\section{Real-Time Subtyping Machines}
\label{section:real}
\input{real}

\section{Implementation and Performance Experiment}
\label{section:implementation}
\input{implementation}

\newpage

\bibliographystyle{plain}
\bibliography{00}

\end{document}

%% file: abstract.tex
\noindent
Grigore showed that Java generics are Turing complete by describing a reduction from Turing machines to Java subtyping.
We apply Grigore's algorithm to Python type hints and deduce that they are Turing complete.
In addition, we present an alternative reduction in which the Turing machines are simulated in real time, resulting in significantly lower compilation times.
Our work is accompanied by a Python implementation of both reductions that compiles Turing machines into Python subtyping machines.

%% file: aa.tex
Python enhancement proposal (PEP) 484 introduced optional type hints to the Python programming language, together with a complete gradual type system~\cite{pep:484}.
Tools such as Mypy \cite{mypy} use type hints to check whether Python programs are correctly typed.
However, certain programs cause Mypy to enter an infinite loop (we show an example below).
We argue that the reason behind these failures is not a Mypy bug, but a deeper issue in the Python type system.
We use Grigore's reduction from Turing machines (TMs) to nominal subtyping with variance \cite{Grigore:2017} to prove that type hints are, in fact, Turing complete.
In other words, checking whether a Python program is correctly typed is as hard as the halting problem.

\subsection{Nominal Subtyping with Variance}

Subtyping is a type system decision problem.
Given types $t$ and $s$, the type system should decide whether type $t$
is a subtype of $s$, \[
    t \subtype s,
\] meaning that every $t$ object is also a member of $s$.
For example, every string is an object, $\text{\lstinline/str/}\subtype \text{\lstinline/object/}$, but not every object is a string, $\text{\lstinline/str/}\nsubtype \text{\lstinline/object/}$.
Subtyping is found, e.g., in variable assignments:
\begin{lstlisting}[language=python]
x: t = ...
y: s = x  # t$\color{comment}\subtype$s
\end{lstlisting}
The second assignment compiles if and only if $t \subtype s$.

Object-oriented languages such as Scala, C\#, and Python use \emph{nominal subtyping with variance} \cite{Kennedy:2007}.
\emph{Nominal} means that subtyping is guided by inheritance.
Type \inline{c1} is a subtype of type \inline{c2} if class \inline{c1} is a descendant of \inline{c2}:
\begin{lstlisting}[language=python]
class s: ...
class t(s): ...
x: t = ...
y: s = x  # vvv
\end{lstlisting}
\emph{Variance} enables variability in type arguments.
By default, type \inline{t[v]} is a subtype of \inline{t[w]} if and only if $\text{\inline{v}}=\text{\inline{w}}$.
If \inline{t}'s type parameter is covariant, then \inline{v} can be any subtype of \inline{w}, $\text{\inline{v}}\subtype\text{\inline{w}}$;
If it is contravariant, then $\text{\inline{v}}\suptype\text{\inline{w}}$.
In Python, variance is specified in an argument to \inline{TypeVar} (the constructor of type parameters).
For example, the following Python program uses contravariance and is correctly typed:
\begin{lstlisting}
X = TypeVar("X", contravariant=True)
class t(Generic[X]): ...
x: t[int] = t[object]()  # vvv (int$\color{comment}\subtype$object)
\end{lstlisting}

Kennedy and Pierce showed that nominal subtyping with variance is undecidable \cite{Kennedy:2007} using a reduction from the Post correspondence problem (PCP) \cite{Post:1946}.
Their work focused on three subtyping features (characteristics):
\begin{enumerate}
    \item \emph{Contravariance}:
        The presence of contravariant type parameters, as described above.
    \item \emph{Expansively recursive inheritance}:
        The closure of types under the inheritance relation and type decomposition ($\text{\inline{t[v]}} \rightarrow \text{\inline{v}}$) is unbounded.
        (The concept of expansive inheritance was introduced by Viroli using a different definition \cite{Viroli:2000}.)
    \item \emph{Multiple instantiation inheritance}:
        Class \inline{t} is allowed to derive class \inline{s[$\cdot$]} multiple times using different type arguments:
\begin{lstlisting}[language=python]
class t(s[object], s[int]): ...  # not legal in Python
\end{lstlisting}
\end{enumerate}
Kennedy and Pierce proved that subtyping becomes decidable when contravariance or expansive inheritance are removed, but they were uncertain about the contribution of multiple instantiation inheritance to undecidability.

\subsection{Subtyping Machines}

Ten years later, Grigore showed that Java subtyping is Turing complete using a direct reduction from TMs \cite{Grigore:2017}.
This reduction uses a subset of Java that conforms to Kennedy and Pierce's nominal subtyping with variance.
Grigore's result settles Kennedy and Pierce's open problem since Java does not support multiple instantiation inheritance \cite[\S8.1.5]{JLS:2015}.
(Intuitively, multiple instantiation inheritance corresponds to non-deterministic subtyping \cite{Kennedy:2007,Roth:2021}, so while it was useful in the PCP reduction, it is redundant in the TM reduction because deterministic TMs are as expressive as non-deterministic TMs.)

Grigore's reduction can be described as a function
\begin{equation}\label{eq:encoding:function}
    e(M, w) = \Delta \cdot q
\end{equation}
that encodes TM $M$ and input word $w$ as program $P=\Delta \cdot q$ consisting of class table $\Delta$ and subtyping query $q$.
The encoding ensures that TM $M$ accepts $w$ if and only if query $q$ type-checks against $\Delta$.

\begin{figure}[ht]
  \vspace{1ex}
  \centering
  \begin{tikzpicture}
    \tikzstyle{cell}=[draw,minimum width=5ex,minimum height=3.5ex,text=#1]
    \tikzstyle{bcell}=[cell=botcolor]
    \tikzstyle{scell}=[cell=symbcolor]
    \node[text=botcolor] (dots1) {$\cdots$};
    \node[bcell] (b1) [right=0pt of dots1] {$\bot$};
    \node[bcell] (b2) [right=-.5pt of b1] {$\bot$};
    \node[scell] (a1) [right=-.5pt of b2] {$a_1$};
    \node[scell] (a2) [right=-.5pt of a1] {$a_2$};
    \node[text=symbcolor] (dots2) [right=0pt of a2] {$\cdots$};
    \node[scell] (am) [right=0pt of dots2] {$a_m$};
    \node[bcell] (b3) [right=-.5pt of am] {$\bot$};
    \node[bcell] (b4) [right=-.5pt of b3] {$\bot$};
    \node[text=botcolor] (dots3) [right=0pt of b4] {$\cdots$};
    \draw[->,thick,headcolor] ([yshift=-2ex]a1.south)--(a1.south);
    \node[draw,circle,text=statecolor] (state) [below=2ex of a1.south] {$q_I$};
  \end{tikzpicture}
  \caption{The initial configuration of a Turing machine}
  \label{figure:tm:configuration}
\end{figure}
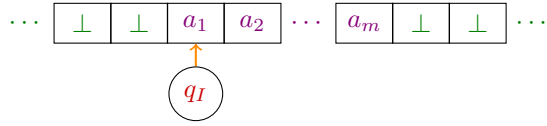

The idea behind the reduction is to encode the TM configuration in the subtyping query $q$.
Recall that the configuration of TM $M$ comprises \1 the content of the memory tape, \2 the location of the machine head on the tape, and \3 the current state of the machine's finite control.
For example, \cref{figure:tm:configuration} illustrates the initial configuration of $M$:
The input word $w=a_1a_2\ldots a_m$ is written on the tape, the machine head points to the first letter $a_1$, and the current state is the initial state $q_I$.

To explain how configurations are encoded as subtyping queries, let us first introduce some syntax (adopted from Grigore's paper).
We write a generic type \inline{A<B<C>>} as $ABC$ for short.
The use of $\fsubtype$ instead of $\subtype$ in a subtyping query means that the type on the left-hand side should be read in reverse (the same goes for $\fsuptype$ and $\suptype$), e.g., $ABC \fsubtype DE$ is equivalent to $CBA \subtype DE$.

The initial TM configuration, depicted in \cref{figure:tm:configuration}, is encoded by the following subtyping query:
\begin{equation}\label{eq:initial:query}
  {\color{botcolor}ZEEL_ \#}N{\color{headcolor}M^L}N{\color{symbcolor}L_{a_1}}N{\color{symbcolor}L_{a_2}}N \cdots
  N{\color{symbcolor}L_{a_m}}N{\color{botcolor}L_ \#}{\color{statecolor}Q_I^{wR}}
  \fsubtype {\color{botcolor}EEZ}
\end{equation}
Observe that the types in \cref{eq:initial:query} have the same colors as the machine configuration elements in \cref{figure:tm:configuration} which they encode.
For example, the type $L_{a_1}$ encodes the tape symbol $a_1$, and both are colored in purple.
As the type on the left-hand side is written in reverse, it is possible to obtain the content of the encoded tape by reading the $L$ types from the left to the right.
The type $EEZ$ at both ends of the query encodes an infinite sequence of blank symbols.
The machine state $q_I$ is encoded by the type $Q^{wR}_I$, and the machine head by the type $M^L$ (the superscripts vary).

Grigore referred to the subtyping query in \cref{eq:initial:query} as a \emph{subtyping machine} because when the subtyping algorithm tries to resolve it, it simulates the computation steps of the original TM.
The state type $Q_I$ is moved along the tape until it reaches the head type $M^L$.
At that point, the subtyping algorithm simulates a single TM transition by overwriting the current tape cell $L_{a_1}$, moving the machine head, and changing the machine state.
The resulting subtyping query correctly encodes the next configuration in the TM run.
This process continues until the machine accepts, and the query is resolved, or the machine aborts and a compilation error is raised.
If the machine runs indefinitely, the subtyping algorithm does not terminate.

While TMs move the machine head to the left or right freely, subtyping machines can change direction only when reaching the end of the tape $EEZ$.
After simulating a TM transition, the subtyping machine must reach the end(s) of the tape, rotate, and then reach the location of the machine head $M$ in the right orientation, before it can simulate the next transition.
In general, Grigore's subtyping machines can make $O(m)$ operations for every computation step of the TM they simulate, where $m$ is the number of symbols on the tape, resulting in a substantial slowdown.
For example, Grigore's simulation of the CYK algorithm, which usually runs in $O(n^3)$, takes $O(n^9)$ subtyping deduction steps to be completed.

%% file: python.tex
Type hints were introduced to the Python programming language with PEP 484 \cite{pep:484}.
Type hints are used as annotations and are entirely optional:
\begin{lstlisting}[language=python]
def positive(x: int) -> bool:
    return x > 0
\end{lstlisting}
Static analysis on type hints is not performed by the Python interpreter, but by third-party tools.
For instance, Mypy \cite{mypy} is a type checker for Python type hints (in fact, PEP 484 was originally inspired by Mypy \cite{pep:484}).

The type system described in PEP 484 supports declaration site nominal subtyping with variance.
Type variables are specified using a special constructor \inline{TypeVar} that can be set as covariant or contravariant.
For example, class \inline{N} in \cref{lst:infinite:subtyping} has a contravariant type parameter \inline{Z}.

As PEP 484 and related PEPs do not put any restrictions on expansively recursive inheritance (at least, that the author could find), we conclude that Python implicitly supports expansive inheritance.
For example, class \inline{C} in \cref{lst:infinite:subtyping} has expansively recursive inheritance, and it correctly compiles in Python and Mypy (the type \inline{"C[C[X]]"} is in a string literal because it forward-references class \inline{C}).

By enabling both contravariance and expansively recursive inheritance, the designers of PEP 484 have subjected Python type hints to the same pitfalls of nominal subtyping with variance studied by Kennedy, Pierce, and Grigore.
For example, the code in \cref{lst:infinite:subtyping}, adapted from Kennedy and Pierce \cite{Kennedy:2007}, shows how contravariance and expansive inheritance can be combined to induce an infinite subtyping cycle.

\begin{lstlisting}[float=ht,caption={Contravariance, expansively recursive inheritance, and infinite subtyping with Python type hints},label={lst:infinite:subtyping}]
from typing import TypeVar, Generic, Any
Z = TypeVar("Z", contravariant=True)
class N(Generic[Z]): ...
X = TypeVar("X")
class C(Generic[X], N[N["C[C[X]]"]]): ...
_: N[C[Any]] = C[Any]() # infinite subtyping
\end{lstlisting}

Contravariance and expansive inheritance are enough to implement Grigore's subtyping machines, which means that Python type hints are Turing complete.
In \cref{section:implementation} we present an implementation of Grigore's reduction with Python type hints.
The resulting subtyping machines correctly run with Mypy.

Grigore explains that subtyping machines could be used to perform certain metaprogramming tasks, and even be abused to attack the compiler \cite{Grigore:2017}.
Most of these (ab)uses still apply to Python type hints with (external) type checking, as provided by Mypy.

Kennedy and Pierce proved that nominal subtyping with variance becomes decidable once expansive inheritance is removed \cite{Kennedy:2007}.
Restrictions to expansive inheritance are implemented in, e.g., the Scala programming language \cite[\S{}5.1.5]{sls} and the .NET framework \cite[\S{}II.9.2]{cli:2005}.
See Greenman et al.~\cite{Greenman:2014} for a further discussion on expansively recursive inheritance and how to deal with it.
Removing expansive inheritance, however, might not be enough to make Python type hints decidable---a more thorough inspection of the various type system features in PEP 484 is required to determine this.
Evidently, any modification to PEP 484 aimed at making it decidable would technically break backward compatibility.

%% file: real.tex
Grigore's subtyping machines must scan the entire tape memory before they can simulate a single TM transition.
This is because the subtyping machines can change their direction (from $\fsubtype$ to $\fsuptype$ and vice versa) only when reaching the end of the tape.
We now present an alternative design for subtyping machines that simulate TMs in real time, i.e., where a single TM computation step is simulated by $O(1)$ subtyping deductions.

Let $M=\langle Q, \Sigma, q_I, q_h, \delta \rangle$ be a TM, where $Q$ is the set of machine states, $\Sigma$ is the set of tape symbols, $q_I$ is the initial state, $q_h$ is the termination state, and \[
    \delta: Q \times \Sigma \rightarrow Q \times \Sigma \times \{L,R\}
\] is the transition function.
The machine head can be moved to the left ($L$) or right ($R$).
The TM accepts its input if and only if it reaches the termination state $q_h$.
We use the symbols $\bot,\# \not \in \Sigma$ to denote blank tape cells.

The new encoding of subtyping machines is shown in \cref{table:new:encoding}.
The encoding comprises ten inheritance rules \1 to \ii{10} (we use a colon to denote inheritance).
To encode a TM $M$ as a subtyping machine, complete the inheritance rules using elements from $M$ that satisfy the conditions on the right-hand side.
For example, if $M$ contains state $q_4 \in Q$ and tape symbol $p \in \Sigma$, then by rule \5 we get the inheritance rule \[
    Q^{LL}_4x:L_pNQ^L_4L_pNx.
\]
Symbol $x$ is a type parameter and symbol \texttt{?}~denotes the wildcard type, i.e., the type that is consistent with (can be substituted by) any type.
In Python, this is the \inline{Any} type \cite{pep:484}.
All the type parameters used in the encoding are contravariant.
There are ten more inheritance rules in addition to those in \cref{table:new:encoding}, obtained by swapping $L$ and $R$ in each rule in the table.

\begin{table}[ht]
  \begin{adjustbox}{center}
    \boxed{
      \renewcommand{\arraystretch}{1.2}
    \begin{tabular}{l}
      $\begin{array}{llr}
        \1~& Q_s^L x: L_a N Q_{s'}^L L_b N x &\delta(q_s,a)=\langle q_{s'}, b, L\rangle \\
        \2~& Q_s^L x: L_a Q_{s'}^{LRR} N L_b N x &\delta(q_s,a)=\langle q_{s'}, b, R\rangle \\
        \3~& Q_s^L x: L_ \# Q_{s'}^{L \#L} N L_b N x &\delta(q_s,\bot)=\langle q_{s'}, b, L\rangle \\
        \4~& Q_s^L x: L_ \# Q_{s'}^{L \#R} N L_b N x &\delta(q_s,\bot)=\langle q_{s'}, b, R\rangle \\
        \5~& Q_s^{LL}x: L_a N Q_s^L L_a N x &\forall q_s\in Q,\forall a\in\Sigma \\
        \6~& N x: Q_s^{LRR} N Q_s^{RR} x &\forall q_s\in Q  \\
        \7~& N x: Q_s^{L \#L} Q_s^{RL} N L_ \# N x &\forall q_s\in Q  \\
        \8~& N x: Q_s^{L \#R} N Q_s^{RR} L_ \# N x &\forall q_s\in Q  \\
        \9~& N x: Q_s^{LR} N Q_s^R x &\forall q_s\in Q  \\
        \ii{10} & Q_h^Lx: L_a N\cc? &\forall a\in\Sigma\cup\{\#\}  \\
      \end{array}$
    \end{tabular}}
  \end{adjustbox}
  \caption{Real-time subtyping machines. For each inheritance rule, swap $L$ and $R$ to get the symmetrical rule.}
  \label{table:new:encoding}
\end{table}

The roles of the types in \cref{table:new:encoding} are mostly the same as in Grigore's encoding~\cref{eq:initial:query}, i.e., $L_a$ is a tape cell containing the symbol $a$, $N$ is a buffer type, and $Q_s$ is a state type.
But, the new encoding does not use a type for the machine head ($M$ in Grigore's encoding).
This is because the state type $Q$ also indicates the location of the machine head.
The superscripts of $Q$ imply the head's movement direction, e.g., $Q^{LL}_s$ means that the head is about to go two cells to the left, $Q^{L\#R}_s$ means that the head is about to move left into a blank cell and then rotate, and so on.

The initial TM configuration is encoded by the following subtyping query:
\begin{equation}\label{eq:new:initial:query}
  {\color{botcolor}ZNL_\#}\twocol{$Q_I^R$}{statecolor}{headcolor}
  {\fsubtype} {\color{symbcolor}L_{a_1}}N{\color{symbcolor}L_{a_2}}N \cdots
  {\color{symbcolor}L_{a_{m-1}}}N{\color{symbcolor}L_{a_m}}N{\color{botcolor}L_ \#NZ}
\end{equation}
The content of the tape is encoded by the $L$ types, read from the left to the right.
The current state and the position of the machine head are encoded by the type $Q^R_I$---the current cell is on the right ($R$) of the type, which is also the direction of the query ($\fsubtype$).
The infinite blank ends of the tape are encoded by the type $L_\#NZ$.
Observe that the colors of the types in \cref{eq:new:initial:query} match the colors of the corresponding elements in the TM in \cref{figure:tm:configuration}.
Type $Q^R_s$ is colored half red and half orange because it represents both the current state and the location of the head.

To prove the correctness of the simulation, we show that the subtyping query in \cref{eq:new:initial:query} simulates the TM transitions while preserving the encoding of the machine tape, head, and state comprising the TM configuration.
There are three variables to be considered: the initial orientation of the head, whether or not the current cell is blank, and whether or not the machine head changes direction.
For example, the query in \cref{eq:new:initial:query} goes to the right ($R$) and reads a non-blank cell $L_{a_1}$;
The next cell could be either $L_{a_2}$, if the head continues right, or blank $L_\#$, if it rotates.

We now cover four cases, assuming that the initial orientation is \emph{left} (this is the orientation used in \cref{table:new:encoding}).
The other four cases are symmetrical.
Next to each subtyping deduction step, we mention the subtyping rule used in this step.
There are two subtyping rules \cite{Kennedy:2007}:
\textsc{Super}, allowing us to replace a type with its supertype using an inheritance rule from \cref{table:new:encoding}, and \textsc{Var}, allowing us to remove a single type from both sides of the query.
As all type parameters are contravariant, the query changes direction (from $\fsubtype$ to $\fsuptype$ and vice versa) after applying \textsc{Var} a single time.

\paragraph*{Case I}
The head points to a \emph{non-blank} cell $a$, replaces it with symbol $b$, and continues \emph{left}.
The relevant TM transition is \[
    \delta(q_s,a)=\langle q_{s'}, b, L\rangle
\]
and the resulting subtyping query is \[
    \begin{array}{ll}
        \gdots N L_a \fsuptype Q_s^L \gdots & \\
        \gdots N L_a \fsuptype L_a N Q_{s'}^L L_b N \gdots & \1+(\textsc{Super}) \\
        \gdots \fsuptype Q_{s'}^L L_b N \gdots & (\textsc{Var})\times2 \\
    \end{array}
\]
(``$\1+\textsc{Super}$'' means applying rule \textsc{Super} with inheritance rule \1 from \cref{table:new:encoding}.
``$(\textsc{Var})\times2$'' means applying rule \textsc{Var} twice.)
Observe that the subtyping query simulates the TM transition while preserving the encoding: symbol $L_a$ is replaced by $L_b$, state $Q_s$ is replaced by $Q_{s'}$ (the next machine state), and the head moves to the cell on the left.

\paragraph*{Case II}
The head points to a \emph{non-blank} cell $a$, replaces it with symbol $b$, and continues \emph{right}.
The relevant TM transition is \[
    \delta(q_s,a)=\langle q_{s'}, b, R\rangle
\]
and the resulting subtyping query is \[
      \begin{array}{ll}
        \gdots N L_a \fsuptype Q_s^L \gdots & \\
        \gdots N L_a \fsuptype L_a Q_{s'}^{LRR} N L_b N \gdots & \2+(\textsc{Super}) \\
        \gdots N \fsubtype Q_{s'}^{LRR} N L_b N \gdots & (\textsc{Var}) \\
        \gdots Q_{s'}^{RR} N Q_{s'}^{LRR} \fsubtype Q_{s'}^{LRR} N L_b N \gdots & \6~+(\textsc{Super}) \\
        \gdots Q_{s'}^{RR} \fsubtype L_b N \gdots & (\textsc{Var})\times2 \\
        \gdots N L_b Q_{s'}^R N L_b \fsubtype L_b N \gdots & \5~+(\textsc{Super}) \\
        \gdots N L_b Q_{s'}^R \fsubtype \gdots & (\textsc{Var})\times2 \\
      \end{array}
\]

\paragraph*{Case III}
The head points to a \emph{blank} cell (the end of the tape), replaces it with symbol $b$, and continues \emph{left}.
The relevant TM transition is \[
    \delta(q_s,\bot)=\langle q_{s'}, b, L\rangle
\]
and the resulting subtyping query is \[
      \begin{array}{ll}
        Z N L_\# \fsuptype Q_s^L \gdots & \\
        Z N L_\# \fsuptype L_\# Q_{s'}^{L\#L} N L_b N \gdots & \3~+(\textsc{Super}) \\
        Z N \fsubtype Q_{s'}^{L\#L} N L_b N \gdots & (\textsc{Var}) \\
        Z N L_\# N Q_{s'}^{RL} Q_{s'}^{L\#L} \fsubtype Q_{s'}^{L\#L} N L_b N \gdots & \7~+(\textsc{Super}) \\
        Z N L_\# N Q_{s'}^{RL} \fsuptype N L_b N \gdots & (\textsc{Var}) \\
        Z N L_\# N Q_{s'}^{RL} \fsuptype Q_{s'}^{RL} N Q_{s'}^L L_b N \gdots & \9~+(\textsc{Super}) \\
        Z N L_\# \fsuptype Q_{s'}^L L_b N \gdots & (\textsc{Var})\times2 \\
      \end{array}
\]

\paragraph*{Case IV}
The head points to a \emph{blank} cell, replaces it with symbol $b$, and continues \emph{right}.
The relevant TM transition is \[
    \delta(q_s,\bot)=\langle q_{s'}, b, R\rangle
\]
and the resulting subtyping query is \[
      \begin{array}{ll}
        Z N L_\# \fsuptype Q_s^L \gdots & \\
        Z N L_\# \fsuptype L_\# Q_{s'}^{L\#R} N L_b N \gdots & \4~+(\textsc{Super}) \\
        Z N \fsubtype Q_{s'}^{L\#R} N L_b N \gdots & (\textsc{Var}) \\
        Z N L_\# Q_{s'}^{RR} N Q_{s'}^{L\#R} \fsubtype Q_{s'}^{L\#R} N L_b N \gdots & \8~+(\textsc{Super}) \\
        Z N L_\# Q_{s'}^{RR} \fsubtype L_b N \gdots & (\textsc{Var})\times2 \\
        Z N L_\# N L_b Q_{s'}^R N L_b \fsubtype L_b N \gdots & \5~+(\textsc{Super}) \\
        Z N L_\# N L_b Q_{s'}^R \fsubtype \gdots & (\textsc{Var})\times2 \\
      \end{array}
\]

The TM rejects its input when its current state is $q_s$, the current tape symbol is $a$, and transition $\delta(q_s, a)$ is not defined.
In this case, the subtyping query $L_a \fsuptype Q^L_s$ also rejects, because there is no inheritance rule in \cref{table:new:encoding} with which rule \textsc{Super} can be applied.
On the other hand, if the TM reaches state $q_h$ and accepts its input, then the subtyping query $L_a \fsuptype Q^L_h$ is resolved by applying rule \ii{10}.

Note that the new simulation is clearly real-time.
To simulate a single TM transition, the subtyping machine performs at most eight subtyping deductions (in cases II and IV).

The wildcard type used in rule \ii{10} does not appear in Kennedy and Pierce's system of nominal subtyping with variance.
Instead of using the wildcard, the subtyping machine could go to either side of the tape before resolving the query, as done in Grigore's simulation.
This makes the design of the subtyping machines more complicated, and their simulation of the TM returns to be non-real-time---although it does not increase the computational complexity of the simulated TM.

%% file: implementation.tex
This paper is accompanied by a Python implementation \cite{Roth:2022} of Grigore's original reduction and our new real-time simulation, introduced in \cref{section:real}.
Our implementation compiles TMs into Python subtyping machines that use the type hints and generics described in PEP 484 \cite{pep:484}.
Each subtyping machine comprises a class table (\cref{table:new:encoding}) and a variable assignment that invokes a subtyping query \cref{eq:new:initial:query}.
To run the subtyping machine, we run Mypy on the generated Python code.

Mypy's subtyping algorithm is most likely implemented using recursion.
As evidence, when running Mypy on the infinite subtyping query described in \cref{lst:infinite:subtyping}, it does not run indefinitely but instead raises a segmentation fault.
This observation makes it possible to measure the run time of the subtyping machine, i.e., the number of subtyping deductions it performs, by the minimal size of the call stack required by Mypy to type-check the machine.

\begin{figure}[ht]
  \centering
  \small
  \begin{tikzpicture}
    \begin{axis}[
      xlabel={Input length},
      ylabel={Call stack size [MB]},
      xmin=5, xmax=95,
      ymin=4, ymax=14,
      xtick={10, 20, 30, 40, 50, 60, 70, 80, 90},
      ytick={5, 7, 9, 11, 13},
      legend pos=north east,
    ]
    \addplot[
      color=blue,
      mark=o,
    ] coordinates {
      (10,5)(12,7)(14,9)(16,13)
    };
    \addlegendentry{Grigore \cite{Grigore:2017}}
    \addplot[
      color=Magenta,
      mark=x,
    ] coordinates {
      (10,5)(20,5)(30,5)(40,5)(50,5)(60,5)(70,7)(80,9)(90,9)
    };
    \addlegendentry{Current work}
    \end{axis}
  \end{tikzpicture}
  \caption{Run times of the palindromes subtyping machines: Grigore's
  reduction vs.~the new reduction.}
  \label{figure:Turing:graphs}
\end{figure}
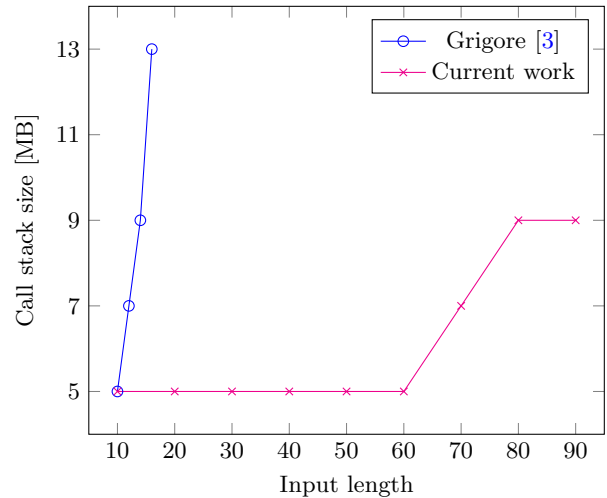

\Cref{figure:Turing:graphs} describes the results of our experiment \cite{Roth:2022}, in which we measured the run times of subtyping machines accepting input words of various lengths.
The TM used in the experiment recognizes palindromes over $\{a,b\}$ and runs in $O(n^2)$.
We compiled the TM together with random palindromes of various lengths into subtyping machines, once using Grigore's method and once with our construction.
Then, we used binary search to find the minimal call stack size (in megabytes (MB)) required for Mypy to type-check the machine without getting a segmentation fault.
The results of the experiment are shown in \cref{figure:Turing:graphs}.

In theory, Grigore's subtyping machines should run in $O(n^3)$ due to their inherent slowdown, while our machines are expected to run in $O(n^2)$ since they simulate the palindromes TM in real time.
In practice, we see that our subtyping machines are much faster than Grigore's and require much fewer resources to be type-checked.